\documentclass{scienceadvhack}

\usepackage{scicite}
\usepackage{times}
\usepackage{graphicx}
\usepackage{amsmath}
\usepackage{amssymb}
\usepackage{url}
\usepackage{hyperref}

\usepackage{newfloat}
\DeclareFloatingEnvironment[name=Extended Figure]{extFigure}

\newcommand{\tess}{{\it TESS}}

\newcounter{lastnote}

\title{Localised thermonuclear bursts from accreting magnetic white dwarfs}

\author
{Simone Scaringi$^{1\ast}$,
Paul J. Groot$^{2,3,4}$, 
Christian Knigge$^{5}$,
Anthony J. Bird$^{5}$,
E. Breedt$^{6}$,
David A.H. Buckley$^{3,4,7}$,
Yuri Cavecchi$^{8}$,
Nathalie D. Degenaar$^{9}$,
Domitilla de Martino$^{10}$,
Chris Done$^{1}$,
Matteo Fratta$^{1}$,
Krystian Ilkiewicz$^{1}$,
Elmar Koerding$^{2}$,
Jean-Pierre Lasota$^{11,12}$,
Colin Littlefield$^{13,14}$,
Carlo F. Manara$^{15}$,
Mairi O'Brien$^{1}$,
Paula Szkody$^{14}$,
Frank X. Timmes$^{16,17}$. 
\\
\\
\normalsize{$^{1}$Centre for Extragalactic Astronomy, Department of Physics, Durham University, DH1 3LE, UK}\\
\normalsize{$^{2}$Department of Astrophysics/IMAPP, Radboud University, P.O. 9010, 6500 GL, Nĳmegen, The Netherlands}\\
\normalsize{$^{3}$South African Astronomical Observatory,  PO Box 9, Observatory, 7935, Cape Town, South Africa}\\
\normalsize{$^{4}$Department of Astronomy, University of Cape Town, Private Bag X3, Rondebosch, 7701, South Africa}\\
\normalsize{$^{5}$School of Physics and Astronomy, University of Southampton, Highfield, Southampton SO17 1BJ, UK}\\
\normalsize{$^{6}$Institute of Astronomy, University of Cambridge, Madingley Road, Cambridge CB3 0HA, UK}\\
\normalsize{$^{7}$Department of Physics, University of the Free State, PO Box 339, Bloemfontein, 9300, South Africa}\\
\normalsize{$^{8}$Instituto de Astronomia, Universidad Nacional Autonoma de Mexico, Ciudad de Mexico, CDMX 04510, Mexico}\\
\normalsize{$^{9}$Anton Pannekoek Institute for Astronomy, University of Amsterdam, Science Park 904, NL-1098 XH Amsterdam, the Netherlands}\\
\normalsize{$^{10}$INAF-Osservatorio Astronomico di Capodimonte, Salita Moiariello 16, I-80131 Naples, Italy}\\
\normalsize{$^{11}$Nicolaus Copernicus Astronomical Center, Polish Academy of Sciences, ul. Bartycka 18, 00-716 Warsaw, Poland}\\
\normalsize{$^{12}$Institut d’Astrophysique de Paris, CNRS et Sorbonne Universites, UMR 7095, 98bis Boulevard Arago, 75014 Paris, France}\\
\normalsize{$^{13}$Department of Physics, University of Notre Dame, Notre Dame, IN 46556, USA}\\
\normalsize{$^{14}$Department of Astronomy, University of Washington, Seattle, WA 98195, USA}\\
\normalsize{$^{15}$European Southern Observatory, Karl-Schwarzschild-Str. 2, D-85748 Garching, Germany}\\
\normalsize{$^{16}$School of Earth and Space Exploration, Arizona State University, Tempe, AZ 85287, USA}\\
\normalsize{$^{17}$Joint Institute for Nuclear Astrophysics - Center for the Evolution of the Elements, USA}\\
\\
\normalsize{$^\ast$Corresponding author. E-mail: simone.scaringi@durham.ac.uk}
\\
\normalsize{Submitted on 4 October 2021. Accepted for publication in \textit{Nature} on 1 February 2022.}
}

\date{}

\begin{document}

\maketitle 

\begin{abstract}
Nova explosions are caused by global thermonuclear runaways triggered in the surface layers of accreting white dwarfs$^{1-3}$. It has been predicted$^{4-6}$ that localised thermonuclear bursts on white dwarfs can also take place, similar to Type I X-ray bursts observed in accreting neutron stars. Unexplained rapid bursts from the binary system TV Columbae, in which mass is accreted onto a moderately-strong magnetised white dwarf from a low-mass companion, have been observed on several occasions in the past $\approx40$ years$^{7-11}$. During these bursts the optical/UV luminosity increases by a factor of $>3$ in less than an hour and fades over $\approx10$ hours. Fast outflows have been observed in UV spectral lines$^{7}$, with velocities $>3500$ km s$^{-1}$, comparable to the escape velocity from the white dwarf surface. Here we report on optical bursts observed in TV Columbae as well as in two additional accreting systems, EI Ursae Majoris and ASASSN$-$19bh. The bursts have a total energy $\approx~10^{-6}$ those of classical nova explosions (``micronovae''), and bear a strong resemblance to Type I X-ray bursts$^{12-14}$. We exclude accretion or stellar magnetic reconnection events as their origin and suggest thermonuclear runaway events in magnetically-confined accretion columns as a viable explanation.
\end{abstract}

TV Columbae (hereafter TV Col) has been extensively studied in the past at wavelengths ranging from optical to hard X-rays$^{7-10}$. The orbital period of 5.5 hr, as well as the white dwarf spin period of 1900 sec have been observed at both optical and X-ray wavelengths$^{15-17}$. During the bursts high-ionisation helium and nitrogen lines strengthen and a transient outflow with velocity $>3500$ km s$^{-1}$ is observed at peak luminosity, revealed by P-Cygni profiles in UV spectral lines$^{7}$. EI Ursae Majoris (hereafter EI UMa) is also known to harbour a magnetic white dwarf accreting matter from a companion star, and both orbital and white dwarf spin periods have been identified at 6.4 hr and 746 sec respectively$^{18-20}$. ASASSN$-$19bh was recently discovered as a transient by ASAS-SN$^{21,22}$ on 25 January 2019,  displaying a sudden increase in brightness of at least 2.1 mag. Extended Data Fig. 1 shows the X-Shooter spectrum of ASASSN$-$19bh obtained on 1 October 2021 revealing absorption lines from the donor star as well as hydrogen Balmer and helium lines in emission. Although the emission lines are narrow when compared to some accreting white dwarfs (AWD), the X-Shooter spectrum reveals ASASSN$-$19bh as an AWD with a donor star compatible with a K-type star, similar to the long period AWD$^{24}$ CXOGBS J175553.2$-$281633.

The \textit{Transiting Exoplanet Survey Satellite} (\tess) observed TV Col during Cycle 1 (15 November 2018 - 6 January 2019) at 120-s cadence and during Cycle 3 (19 November 2019 - 13 January 2020) at 20-s cadence. EI UMa was observed during \tess\ Cycle 2 (24 December 2019 - 20 January 2020) at 120-s cadence, and ASASSN-19bh during Cycle 3 (5 July 2020 - 30 July 2020 and 29 April 2021 - 24 June 2021) at 120-s cadence. Figure 1 shows the observed \tess\ Cycle 1 (Sectors 32 and 33) lightcurves of TV Col with three consecutive bursts observed, each lasting 12 hr with a rise time of less than 30 minutes, and each burst separated by 3 days. Figure 2 displays the observed \tess\ lightcurve of EI UMa during Sector 20, showing two rapid and consecutive bursts each lasting $\approx7$ hr and separated by $\approx1$ day. The brightness evolution and temporal properties resemble those observed in TV Col, displaying multiple peaks and troughs. Figure 2 also displays the \tess\ lightcurve of ASASSN$-$19bh observed during Sector 38 showing a precursor followed by a single energetic burst with a rise time of $\approx1.5$ hr during which the luminosity increases by a factor of $\approx25$ and decaying over several days. In both Figure~1 and Figure~2 we have calibrated the observed \tess\ count rates into g-band equivalent luminosities using quasi-simultaneous ground-based g-band ASAS-SN observations$^{16,17}$ and distances inferred from the \textit{Gaia}$^{25,26}$ measured parallaxes (see \textit{Methods}). We use the calibrated lightcurves to infer peak burst luminosities and total energy released by each burst. Without taking into account a bolometric correction we constrain the mean peak luminosity of the bursts to be $1.0\times 10^{34}$ erg s$^{-1}$, $2.5\times 10^{34}$ erg s$^{-1}$ and $3.5\times 10^{34}$ erg s$^{-1}$ for TV Col, EI UMa and ASASSN$-$19bh, respectively. Integrating the burst luminosities, and summing the energies in the consecutive bursts observed in TV Col and EI UMa, yields burst energies of $3.5\times 10^{38}$ erg, $5.2\times 10^{38}$ erg and $1.2\times 10^{39}$ erg respectively for TV Col, EI UMa and ASASSN$-$19bh. The luminosity rise gradient during the burst onset for all observed bursts in all three systems is determined to be in the range of 27-52 mag/day ($1.6$-$5.0$ $\times10^{30}$ erg s$^{-2}$), while the rate of decay is observed in the range 1-3 mag/day (0.8-5.3 $\times 10^{29}$ erg s$^{-2}$). The \tess\ observations not only reveal that the rapid bursts observed in TV Col can sometimes occur in clusters, but also that these bursts, as determined by their temporal and energetic properties, are not limited to just TV Col. Furthermore, the long-term ASAS-SN monitoring of these sources (see Extended Data Fig.~2) establishes these bursts as a recurring phenomenon. 

A few scenarios have been proposed to explain the rapid bursts in TV Col, and by analogy we can attempt to apply these to EI UMa and ASASSN$-$19bh. One prominent model is that the bursts are driven by thermal-viscous instabilities in the disk, similar to so-called dwarf-nova (DN) outbursts observed in a range of subtypes of AWDs$^{27,28}$. However, the much shorter duration of the bursts, as well as their occurrence in closely-spaced clusters, invalidates this scenario$^{11,28}$. Another possibility may be that luminosity variations are induced by magnetically gated flares as observed in some weakly magnetised AWDs$^{29,30}$. For this to happen, the disk would need to reach a specific mass transfer rate, and the bursts would appear as quasi-periodic, making this scenario also unlikely in explaining the observed bursts. Enhanced mass transfer events driven by an instability in the donor stars has also been suggested to explain the rapid bursts$^{9}$. In this scenario a short-lived enhanced mass transfer event would allow the stream of material to overflow the outer accretion disk edge. Our observations of consecutive bursts in TV Col and in EI UMa make this scenario also unlikely, since it would require the donor star to drive several consecutive instabilities that maintain their coherence while travelling through the outer accretion disk. Finally, low-mass stars similar to the donors in TV Col, EI UMa, and ASASSN$-$19bh are known to sporadically release energy through stellar flares via magnetic reconnection events with bulk luminosities emitted in the \tess\ passband$^{31-33}$. However, even the most energetic stellar flares observed to date release 4 orders of magnitude less energy than TV Col and EI UMa, and 5 orders of magnitude less than ASASSN$-$19bh$^{34,35}$. We thus also exclude stellar flares as a candidate explanation for the observed rapid bursts. 

The burst luminosity rise gradient for all observed bursts in all three systems is much faster than those observed in DN outbursts and comparable to those observed in thermonuclear runaway (TNR) events such as classical nova explosions$^{36,37}$. \tess\ also reveals how the time evolution of the individual bursts resembles TNR events occurring on the surface of accreting neutron stars, observed at X-ray wavelengths: Type I X-ray bursts$^{12-14}$. The multi-peaked time evolution and the rapid succession of bursts observed in both TV Col and EI UMa mimic those observed in the X-ray binary neutron star 4U 1636$-$536$^{38-40}$, slowed down by about three orders of magnitude. ASASSN$-$19bh also displays a resemblance to those more energetic single peaked Type-I X-ray bursts$^{12}$, and shares a similar precursor to the Type-I X-ray burst observed in the neutron star X-ray binary SAX J1808.4$-$3658$^{41}$ (see also Extended Data Fig. 3). On close inspection, the first burst observed in TV Col by \tess\ (Figure 1, panel b) also reveals the presence of several precursors during the burst onset. Furthermore, the fast rise and slow decay observed in the \tess\ lightcurves is also analogous to what is observed in Type-I X-ray bursts. All bursts observed by \tess\ have comparable burst energies to those observed in Type-I bursts of $10^{38}$-$10^{39}$ erg. These observables form the basis for exploring the possibility that the rapid bursts observed may have a thermonuclear origin.

Given the short duration and the energies released by the bursts compared to nova explosions, the thermonuclear runaway must be restricted to burning a limited amount of material and confined to a fraction of the white dwarf surface$^{4-6}$. In analogy with classical nova outbursts, in order to ignite hydrogen fuel on a carbon-oxygen white dwarf, the accreted material must reach a critical pressure at the base of the accreted layer of the order of $P_{crit}\approx10^{18}$-$10^{19}$ dyn cm$^{-2}$, where the exact value will depend on the white dwarf mass (and thus also radius$^{42}$), the temperature, as well as the specific mass accretion rate per unit surface area$^{3}$. In non-magnetic systems, this high pressure is generally achieved in AWDs when the spherical shell of accreted material reaches $10^{-4}M_{\odot}$-$10^{-6}M_{\odot}$. The amount of accreted mass required to reach $P_{crit}$ can be substantially reduced if accretion onto the white dwarf is confined to a much smaller fractional surface area$^{4-6}$. If we take the mass-to-energy conversion for hydrogen-to-helium fusion flash in the CNO cycle of $10^{16}$ erg g$^{-1}$ we can infer a lower limit on the mass burned during the bursts in TV Col, EI UMa and ASASSN$-$19bh of $1.8\times 10^{-11} M_{\odot}$, $2.6\times 10^{-11}M_{\odot}$ and $5.8\times 10^{-11} M_{\odot}$ respectively, a factor of $\approx10^{-6}$ lower than in classical novae.

Once triggered, a localised TNR will increase in temperature while burning through most of the freshly accreted layer. This process is also known to drive outflows (e.g. classical novae$^{1-3}$) with velocities comparable to, or higher than, the escape velocity of the AWD, after which the localised area is expected to drop back to the quiescent temperature. The sudden appearance of high ionization HeII lines during the rise of two bursts observed in TV Col$^{7,9}$ is consistent with the presence of a hot ionizing source. The same lines weaken during the burst decay$^9$ and are entirely absent during quiescence$^{7}$, suggesting that the ionizing source fades and disappears. Furthermore, the bursts are observed to become bluer during the rise, which in turn suggests that the system must cool as it decays. Finally, high outflow velocities are observed only at the peak of the bursts$^{7}$. All these observables are consistent with the interpretation that the bursts are in fact TNR events.

The cooling of an expanding photosphere following a TNR event has been observed in Type I X-ray bursts from accreting neutron stars$^{14,38}$, providing a further analogy to what is observed in TV Col. Furthermore, by analogy to Type I X-ray bursts, where very short waiting times are sometimes observed for the bursts$^{14,43,44}$, the temporal evolution and short waiting times between the bursts in TV Col and EI UMa may be related to the energy from the TNR being released in steps. In the phenomenology of Type I X-ray bursts, the multi-peak behaviour and short recurrence times are explained as the result of partial burning at the base of the accreted column. This is thought to be driven by opacity changes due to temperature that drive convection in the column. In turn this partially burns the column mass and then re-ignites it$^{45}$. On the other hand the precursor followed by the bright burst observed in ASASSN$-$19bh may indicate that the ignition at the base of the accretion column acts as a trigger that drives a shock front propagating and igniting the entire accumulated column mass in one energetic event. This mechanism has also been suggested to explain the precursor observed in the Type-1 X-ray burst$^{41}$ of SAX J1808.4$-$3658. In the case of ASASSN$-$19bh, the precursor may thus be related to a shock-breakout, somewhat similar to what is observed in a Type II supernova$^{46}$. Alternatively it may also be that the longer burst duration observed in ASASSN$-$19bh is somewhat analogous to the intermediate-duration and superbursts observed in accreting neutron stars$^{47}$.

With a white dwarf surface magnetic field $B > 10^{6}$ G, as expected for TV Col and EI UMa, the spinning magnetosphere is able to govern the accretion dynamics and funnel material onto a smaller surface area, $A_{col}$, creating an accretion column at the white dwarf magnetic poles$^{48,49}$. If the ram pressure from the infalling stream of material is comparable to the critical pressure required to ignite a TNR, $P_{ram}\approx P_{crit}$, a TNR can be expected. For this to happen, however, the fractional area, $f=\frac{A_{col}}{A_{WD}}$, has to be extremely small, of the order of  $f<10^{-10}$ (see \textit{Methods}) when in general the fractional surface area is expected, and observed$^{48,49}$, to be  $10^{-2}>f>10^{-7}$. It is plausible that transverse temperature gradients and inhomogeneities in the accreted layers thermalise on timescales that are much longer than the thermonuclear runaway timescale. This prediction$^{4,6}$ is also expected to drive a localised TNR rather than a spherically symmetric global eruption. 

The identification and characterisation of rapid bursts in three AWDs have unveiled that magnetically confined TNRs may occur in the surface layers of white dwarf atmospheres. Similar bursts to those reported here have also been reported in the literature for other magnetic AWDs (e.g. V1223  Sagittarii$^{9}$ and DW Cancri$^{50}$) and it may be that localised TNRs on AWDs are more common than previously thought. Further observations of similar bursts across the electromagnetic spectrum, as well as detailed theoretical modelling of localised TNRs, are necessary to determine what truly triggers these events.

\begin{figure*}[ht]
\begin{center}
\includegraphics[width = \textwidth]{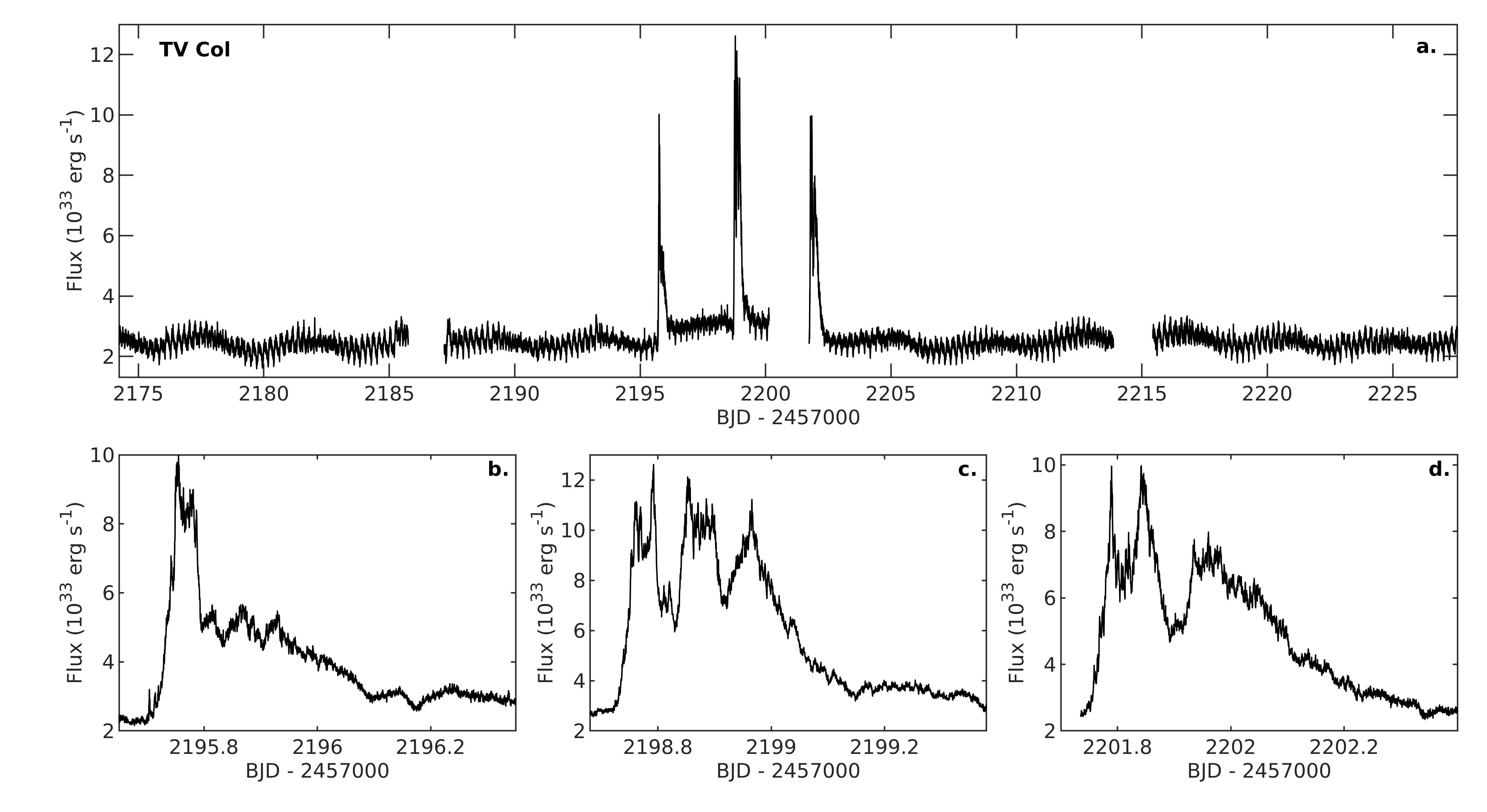}
\caption{
\noindent \textbf{Optical brightness variations in TV Col}
\textbf{a.} \tess\ lightcurve (20-s cadence) of TV Col. The lightcurve has been calibrated against quasi-simultaneous ground based ASAS-SN g-band data (see \textit{Methods}). Panels \textbf{b.}, \textbf{c.} and \textbf{d.} show 16.8 hours of data around the three detected rapid bursts. The individual bursts yield integrated energies of $0.9\times10^{38}$erg, $1.6\times10^{38}$erg and $1.0\times10^{38}$erg, respectively.
}
\end{center}
\end{figure*}

\begin{figure*}[ht]
\begin{center}
\includegraphics[width = \textwidth]{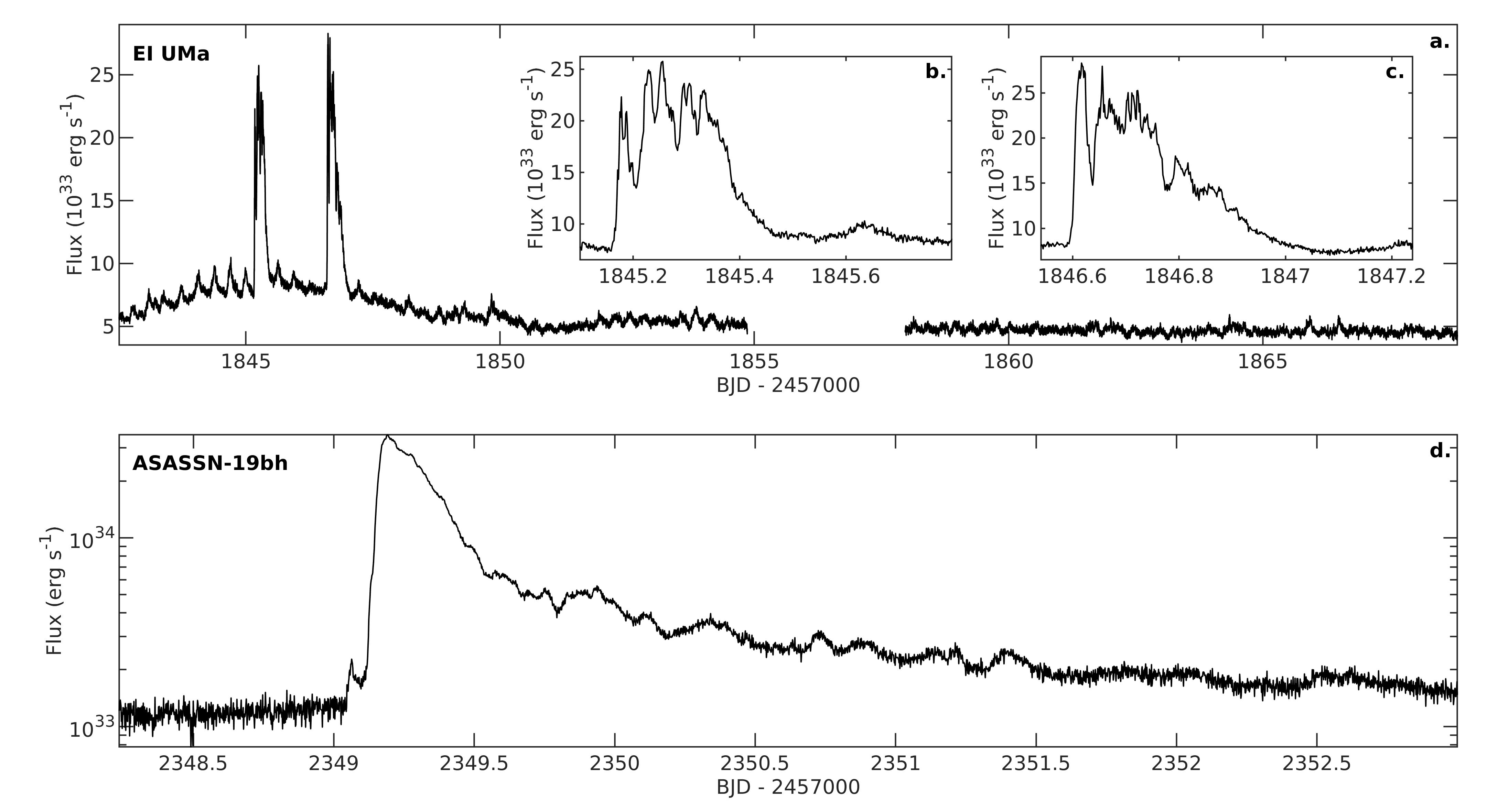}
\caption{
\noindent \textbf{Optical brightness variations in EI UMa and ASASSN$-$19bh.}
\textbf{a.} \tess\ lightcurve (2-min cadence) of EI UMa. Panels \textbf{b.} and \textbf{c.} show 16.8 hours of data around the two detected rapid bursts. The individual bursts yield integrated energies of $2.4\times10^{38}$erg and $2.8\times10^{38}$erg respectively. \textbf{d.} \tess\ lightcurve (2-min cadence) of ASASSN$-$19bh showing the observed single energetic burst of $1.2\times10^{39}$erg. All lightcurves have been calibrated against quasi-simultaneous ground based ASAS-SN g-band data (see \textit{Methods}).
}
\end{center}
\end{figure*}

\section*{METHODS}

\subsubsection*{X-Shooter spectrum of ASASSN$-$19bh}
The Very Large Telescope (VLT) observed ASASSN$-$19bh on 1 October 2021 with X-Shooter51. Exposure times were set at 790s for both the UBV and VIS arm of the spectrograph, and 3 100s for the NIR arm. The data reduction was carried out with the ESO Reflex pipeline$^{52}$ v.3.5.0. The pipeline performs the standard reduction, including flat fielding, bias and dark corrections, wavelength calibration and rectification of the spectrum and extraction. The flux calibration was performed using a standard star observed on the same night. No telluric correction was performed. Extended Data Fig. 1 shows the resulting spectrum in the relevant wavelength ranges 3900\AA-4900\AA and 5800\AA-6700\AA. 

\subsubsection*{Data Sources}
The \tess\ data for TV Col, EI UMa and ASASSN$-$19bh were obtained from The Barbara A. Mikulski Archive for Space Telescopes (MAST) in reduced and calibrated format. The \tess\ telescope/detector is sensitive to light across a wide range of wavelengths (600 nm - 1000 nm). For all, we also retrieved ASAS-SN$^{22,23}$ ground based V- and g-band available photometry. More specifically, this data set includes g-band observations for all systems that are quasi-simultaneous with the \tess\ data. We used this overlap to establish an approximate transformation of the \tess\ count rates for TV Col, EI UMa, and ASASSN-19bh into standard g-band fluxes. To achieve this, we first selected the data points between \tess\ and ASAS-SN that were taken within 120s of each other. These selected data points were correlated and linearly fitted. The fits were performed independently on each half of a \tess\ sector in order to minimise systematic count rate offsets induced between different observation sectors and in-between data downlinks. The resulting linear fits allow the \tess\ count rates to be converted into equivalent ASAS-SN g-band fluxes, but we note that this method does not take into account any bolometric correction. We then adopt the \textit{Gaia} parallax measurements$^{25,26}$ for TV Col, EI UMa and ASASSN-19bh to convert the fluxes into luminosities as shown in Figure~1 and Figure~2. To compute the peak luminosities of each burst we have subtracted an estimate of the quiescent luminosity level. This was measured as the mean luminosity throughout the 1 day preceding each burst. The burst energies have also been estimated using the same quiescent luminosity subtraction. The long term ASAS-SN lightcurves, as well as the \tess\ calibrated photometry, are shown in Extended Data Fig.~2. 

To better compare the bursts observed by \tess\ to those observed at X-ray wavelengths in Type-I X-ray bursts we have retrieved archival data on 4U 1636$-$536 and SAX J1808.4$-$3658. Extended Data Fig.~3 shows insets of bursts from each system compared to the TV Col and ASASSN-19bh lightcurves observed by TESS. The data for 4U 1636$-$536 is from an \textit{EXOSAT}-ME observation$^{39}$ performed on 09-08-1985. The SAX J1808.4$-$3658 is from an \textit{RXTE}-PCA observation$^{41}$ performed on 19-10-2002.

\subsubsection*{Timing analysis of \tess\ data}
TV Col is known to display evidence of both negative and positive superhumps$^{53}$. The former signal suggests the presence of a tilted and precessing accretion disk, while the latter is related to a tidally deformed accretion disk which can occur when the outer disk edge extends to the 3:1 resonance radius with the donor star$^{54}$. When this happens, the positive superhump excess ($\epsilon=\frac{P_{sh}-P_{orb}}{P_{orb}}$, where $P_{sh}$ and $P_{orb}$ are the positive superhump and orbital period respectively) can be used to infer the binary mass ratio. In the case of TV Col, the inferred mass ratio is large ($q\approx0.92$) compared to other AWDs$^{54}$. No positive or negative superhump periods have been reported in the literature for either EI UMa or ASASSN-19bh. 

Lomb-Scargle periodograms$^{55,56}$ of TV Col were computed using TESS Cycle 1 and Cycle 3 data. The periodogram for Cycle 1 data was computed using the available 120-s cadence data, while we used the available 20-s cadence data for Cycle 3. As the 3 bursts observed during Cycle 3 introduce excess power at low frequencies due to red-noise leakage, we excluded the Cycle 3 data in the time range $2195.5<$TJD$<2202.5$, where TJD is the \tess\ Julian date (BJD$-2457000$). Extended Data Fig.~4 shows the computed normalised periodograms. Cycle 1 data displays strong signals at the orbital frequency of $f_{orb}=4.374$(19) c d$^{-1}$ (5.487 hr) as well as several associated harmonic frequencies, consistent with reported literature values$^{8,9,54,57}$. The error on the signal frequency has been inferred through the window function ($\frac{1}{T}$ where $T$ is the length of the lightcurve segment). Computing the signal-to-noise ratio (S/N) of the detected signals requires prior knowledge of the shape of the underlying intrinsic broad-band noise components. In the absence of this we use the root-mean-square of powers at frequencies just short and just long of the detected signal to obtain S/N$>$41 for the orbital signal. A further signal at $f=40.879$(19) c d$^{-1}$ (2114 s) is also detected with a S/N$>$5 using the same methodology as for the orbital signal. This latter signal is consistent with the beat signal between the previously published white dwarf spin signal of $f_{spin}=45.224$ c d$^{-1}$  (1911s)$^{57,58}$ and the system orbital period. The Cycle 1 periodogram also displays strong signals at the system orbital frequency and related harmonics. We also detect a superorbital frequency at $f_{so}=0.257$ c d$^{-1}$ with S/N$>$10 as well as the associated negative superhumps resulting from the superorbital-to-orbital beat frequency at $f_{-}=4.630$(19) c d$^{-1}$ with S/N$>$40. 

Periodograms were computed using the \tess\ 120-s cadence data of EI UMa. Extended Data Fig.~5 shows the periodogram obtained before the bursts commenced (TJD$<$1845.1) and after the bursts occurred (TJD$>$1856). During the pre-burst stage we detect signatures of a positive superhump at $f_{+}=3.08$(39) c d$^{-1}$ with S/N$>$6. During the post-burst phase we detect the system orbital period $f_{orb}=3.729$(92) c d$^{-1}$ (6.435 hr) with S/N$>$7 as well as the spin-to-orbital beat frequency $f=112.240$(92) c d$^{-1}$ (769.8 s) with S/N$>$4, both consistent with values reported in the literature$^{20,21}$. We use the detected positive superhump observed before the burst to infer the period excess of EI UMa to be $\epsilon=0.208$. To date this constitutes the largest period excess detected in an AWD, surpassing that of TV Col$^{53}$. We note however that the positive superhump-to-mass ratio relation has not been calibrated for $\epsilon>0.1$. Furthermore, large mass-ratio binaries are not necessarily expected to follow the same superhump relation as low mass-ratio binaries. It is interesting to note however that the positive superhump signal in EI UMa disappears after the two bursts. This may be due to the localised TNRs disrupting the disk geometry, thus dissipating and eventually quenching the positive superhump signal. A similar mechanism is also invoked in explaining some of the variability observed immediately following intermediate duration Type~I X-ray bursts$^{59,60}$. 

Periodograms of ASASSN-19bh using 120-s cadence data did not reveal any clear coherent significant signals. However, a sinusoidal signal with periodicity of 4.5 days and 20\% amplitude is observed throughout all the \tess\ observations prior to the flare onset. This signal is reminiscent of what is also observed in TV Col during quiescence, and is possibly related to a tilted retrogradely precessing accretion disk as observed in other AWDs$^{61}$. 

We note that while both TV Col and EI UMa are known to be magnetic AWDs from the direct detection of their spin periods and hard X-ray emission$^{62-64}$, the magnetic nature of the white dwarf in ASASSN-19bh is not known. If it were magnetic, the larger distance to ASASSN-19bh of 1.5kpc compared to TV Col (514pc) and EI UMa (1.14kpc) would make the X-ray emission fainter, plausibly explaining the current non-detection in the X-ray band. Future observations will reveal whether ASASSN-19bh is a magnetic AWD, strengthening the evidence for magnetically confined and localised TNRs.

\subsubsection*{Localised TNR from ballistic impact}
When the white dwarf surface magnetic field is $B >10^{6}$ G, the spinning magnetosphere is able to govern the accretion dynamics and funnel material onto a smaller surface fractional area creating an accretion column onto the white dwarf magnetic poles$^{50,65,66}$. In practice the disk truncation radius will depend on the combination of white dwarf spin, mass transfer rate through the disk, and white dwarf surface magnetic field. If the spinning magnetospheric barrier resides within the disk circularisation radius then an accretion disk can form and be truncated at the inner-disk edges. If, on the other hand, the magnetospheric barrier resides at radii larger than the disk circularisation radius, then material from the donor star is able to latch onto field lines before it is able to form a disk. In the most extreme cases, when the white dwarf surface magnetic field is $B >10^{7}$G, the white dwarf spin and binary orbital period become synchronized and mass transfer proceeds directly from the donor star onto the white dwarf magnetic poles along magnetic field lines. The infalling material will exert a ram pressure on the white dwarf magnetic poles from the ballistic impact, $P_{ram} \propto \dot{M} f^{-1}M_{WD}^{1/2}R_{WD}^{-5/2}$, where $\dot{M}$ is the mass accretion rate onto the surface, and $M_{WD}$ and $R_{WD}$ are the white dwarf mass and radius respectively$^{49}$. The factor $f$ is the surface fractional area of the impact on the white dwarf, $f = \frac{A_{col}}{A_{WD}}$. If the pressure at the base of the accretion column is comparable to the critical pressure, such that $P_{b}\approx P_{ram} \approx P_{crit}$, then it may be possible to initiate localised fusion simply from the large pressure of the incoming ballistic accretion flow. For this to happen however a very small fractional area of $f<10^{-10}$ is required even to reach the lowest critical pressure of $P_{ram}\approx P_{crit}\approx 10^{18}$ dyn cm$^{-2}$. 

The fractional area onto which the flow impacts the white dwarf surface will depend on how far away the magnetospheric disk truncation is, which in turn depends on the white dwarf’s magnetic field strength and mass accretion rate. The inhomogeneous accretion flow scenario (referred to as the ``bombardment'' scenario in the low mass transfer regime)$^{49,65}$, which has been successful in explaining several observations of strongly magnetic AWDs, envisages three distinct accretion fractional areas. In this model, the accretion flow onto the magnetic poles does not necessarily have to be homogeneous, but may occur (at least occasionally) through higher density parcels of material referred to as ``blobs''. The largest fractional area considered is $f_{zone}$, which is related to the region over which accretion takes place onto the magnetic poles. The smallest fractional area considered, $f_{acc}$, is related to the impact region of the individual blobs of material raining down onto $f_{zone}$. These blobs are expected to be more elongated when latching onto field lines from larger magnetospheric radii, and because of their smaller fractional area and higher density, are also expected to bury themselves several scale heights below the white dwarf photosphere$^{49}$. The fractional area $f_{eff}$ is related to the effective radiative area of the energy released within the white dwarf photosphere from the buried blobs, which have impacted the white dwarf on $f_{acc}$. Thus, in general, the bombardment scenario defines three accretion fractional areas such that $f_{acc}<<f_{eff}<f_{zone}$.

The smallest fractional areas ($f_{acc}$) in magnetic AWDs can occur in so-called polars, where the accretion stream from the secondary can become highly inhomogeneous as it travels along the field lines directly from the donor star. In this most extreme case the model$^{48,49,65}$ can allow for the filamentary blobs to have an $f_{acc}\approx 10^{-7}$ on impact, at least 3 orders of magnitude larger than what is required to reach $P_{ram}\approx P_{crit}$. Furthermore, assuming the rapid bursts are the result of a localised TNR burning through an accreted hydrogen-rich filament, we can use the hydrogen-to-helium conversion of $\approx 10^{16}$erg g$^{-1}$ released during the CNO flash$^{2,3}$ to infer that the bursts would burn a mass in excess of $>1\times10^{-11}M_{\odot}$. This would then imply that the filaments are of comparable mass, which is excessively high for individual accretion events. Finally, because the white dwarfs in both TV Col and EI UMa are known to be asynchronously rotating with respect to their binary orbits, accretion onto the magnetic field lines is expected to be via a disk rather than from a stream directly from the donor. The magnetically confined accretion flow is thus expected to be more homogeneous than that in polars. It is thus highly unlikely that the required fractional area of $f_{acc}<10^{-10}$ can be achieved to reach $P_{ram}\approx P_{crit}$ and ignite a localised TNR.

\bibliography{scibib}

\bibliographystyle{Science}

\vspace{-0.6cm}
\paragraph*{Acknowledgments:}
P.J.G. is supported by NRF SARChI grant 111692. D.A.H.B. acknowledges research support from the South African National Research Foundation. D.dM. acknowledges financial support from the Italian Space Agency (ASI) and National Institute for Astrophysics (INAF) under agreements ASI-INAF I/037/12/0 and ASI-INAF n.2017-14-H.0 and from INAF ‘Sostegno alla ricerca scientifica main streams dell’INAF’, Presidential Decree 43/2018 and from INAF ‘SKA/CTA projects’, Presidential Decree 70/2016 and from PHAROS COST Action N. 16214. C.D and K.I. acknowledge funding from STFC consolidator grant ST/T000244/1. J.-P.L. was supported in part by a grant from the French Space Agency CNES. P.S. acknowledges support from NSF grant AST-1514737. F.X.T. is supported by the National Science Foundation (NSF) under grant No. ACI-1663684 for the MESA Project, and by the NSF under grant No. PHY-1430152 for the Physics Frontier Center Joint Institute for Nuclear Astrophysics Center for the Evolution of the Elements (JINA-CEE). This paper includes data collected by the \tess\ mission. Funding for the \tess\ mission is provided by the NASA's Science Mission Directorate. Some of the data presented in this paper were obtained from the Mikulski Archive for Space Telescopes (MAST). STScI is operated by the Association of Universities for Research in Astronomy, Inc., under NASA contract NAS5-26555. Support for MAST for non-HST data is provided by the NASA Office of Space Science via grant NNX09AF08G and by other grants and contracts. This paper uses data from the ASAS-SN project run by the Ohio State University. We thank the ASAS-SN team for making their data publicly available. This work has also made use of data from the European Space Agency (ESA) mission \textit{Gaia} (https://www.cosmos.esa.int/gaia), processed by the Gaia Data Processing and Analysis Consortium (DPAC, https://www.cosmos.esa.int/web/gaia/dpac/consortium). Funding for the DPAC has been provided by national institutions, in particular the institutions participating in the Gaia Multilateral Agreement. Based on observations collected at the European Southern Observatory under ESO-DDT programme 107.2309.001, for which the authors acknowledge support from the ESO Director-General.   

\vspace{-0.4cm}
\paragraph*{Author Contributions:}
S.S. was PI of the \tess\ proposal to obtain the data, discovered the bursts and performed the ASAS-SN luminosity calibration, co-developed the application of the bombardment model to potentially drive TNRs, and led the interpretation of the bursts. P.J.G. was PI of the X-Shooter proposal to obtain the spectrum of ASASSN-19bh. Y.C. contributed details on the analogy with Type I X-ray bursts, including leading the discussions on their temporal evolution. C.M. reduced the X-Shooter spectrum of ASASSN-19bh. All authors shared ideas, interpreted the results, commented, and edited the manuscript.

\begin{extFigure*}[ht]
\begin{center}
\includegraphics[width = \textwidth]{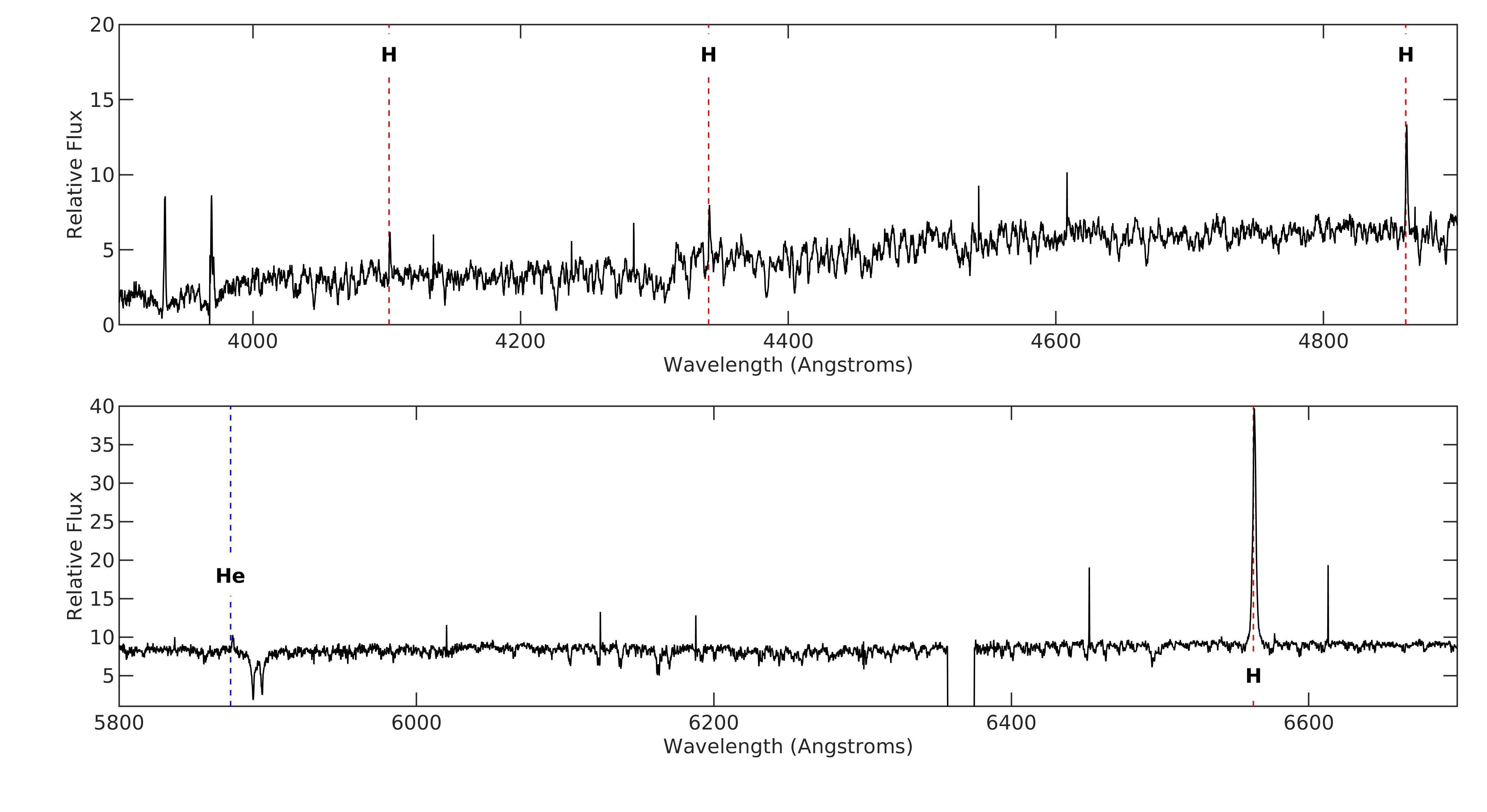}
\caption{
\noindent \textbf{X-Shooter spectrum of ASASSN-19bh.}
\textbf{a.} X-Shooter spectrum (UBV arm) in the range 3900\AA-4900\AA. \textbf{b.} X-Shooter spectrum (VIS arm) in the range 5800\AA-6700\AA. Hydrogen Balmer-series lines are marked with the red dashed lines. The HeI 5875\AA is marked with a blue dashed line. CaII H\&K emission lines can be seen blueward of 4000\AA. Numerous narrow absorption lines from the secondary are also identifiable. Narrow emission spikes in e.g. the region 6000\AA-6200\AA are residuals from the sky subtraction. 
}
\end{center}
\end{extFigure*}

\begin{extFigure*}[ht]
\begin{center}
\includegraphics[width = \textwidth]{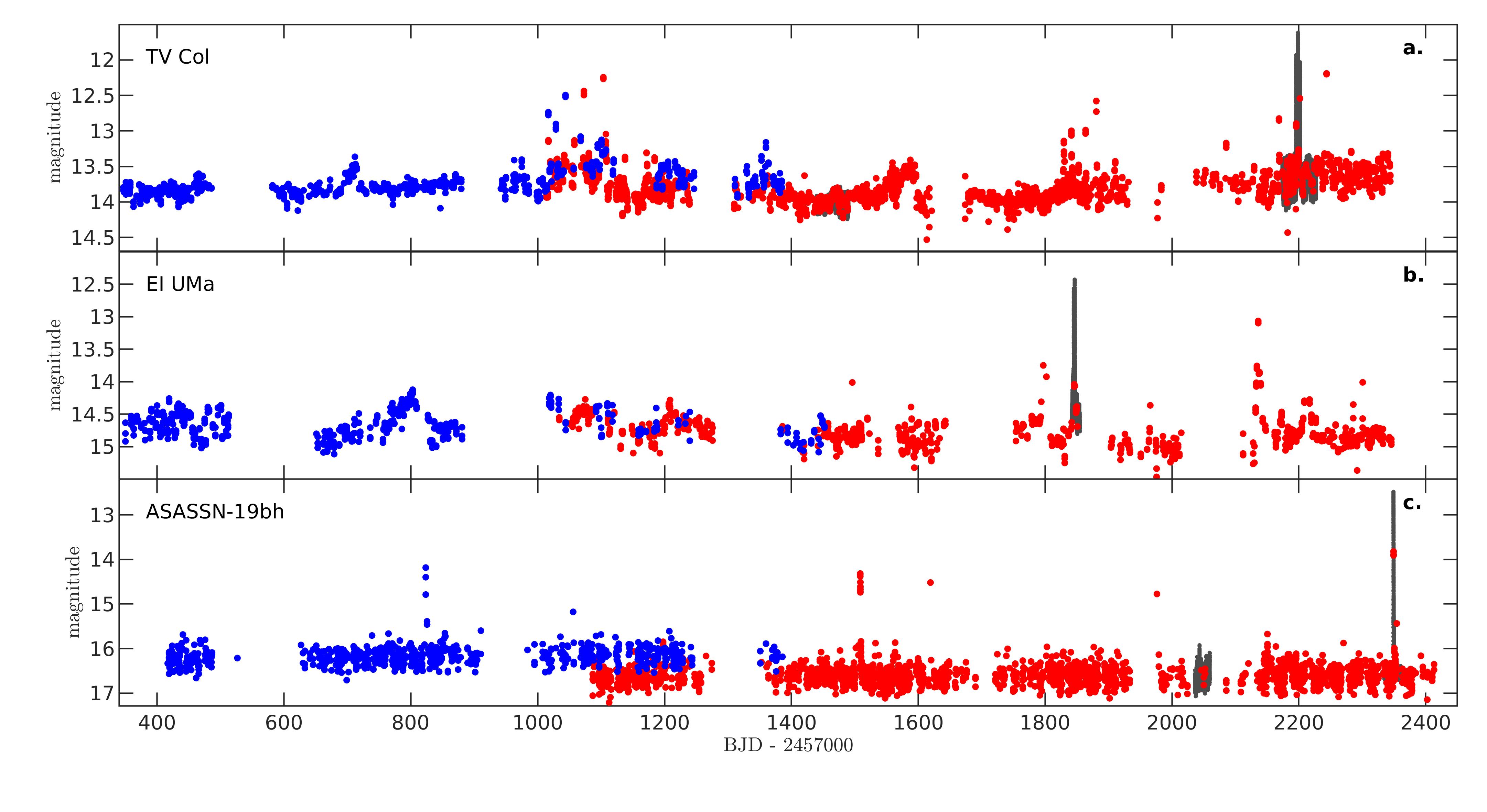}
\caption{
\noindent \textbf{Long term ASAS-SN lightcurves of TV Col, EI UMa, and ASASSN-19bh.}
\textbf{a.} ASAS-SN lightcurve of  TV Col \textbf{b.} ASASSN lightcurve of EI UMa \textbf{c.} ASASSN lightcurve of ASASSN-19bh. In all panels the blue and red points correspond to ASAS-SN V-band and g-band photometry respectively. Calibrated \tess\ data points are shown in grey. Typical uncertainties on magnitude are 0.02.
}
\end{center}
\end{extFigure*}

\begin{extFigure*}[ht]
\begin{center}
\includegraphics[width = \textwidth]{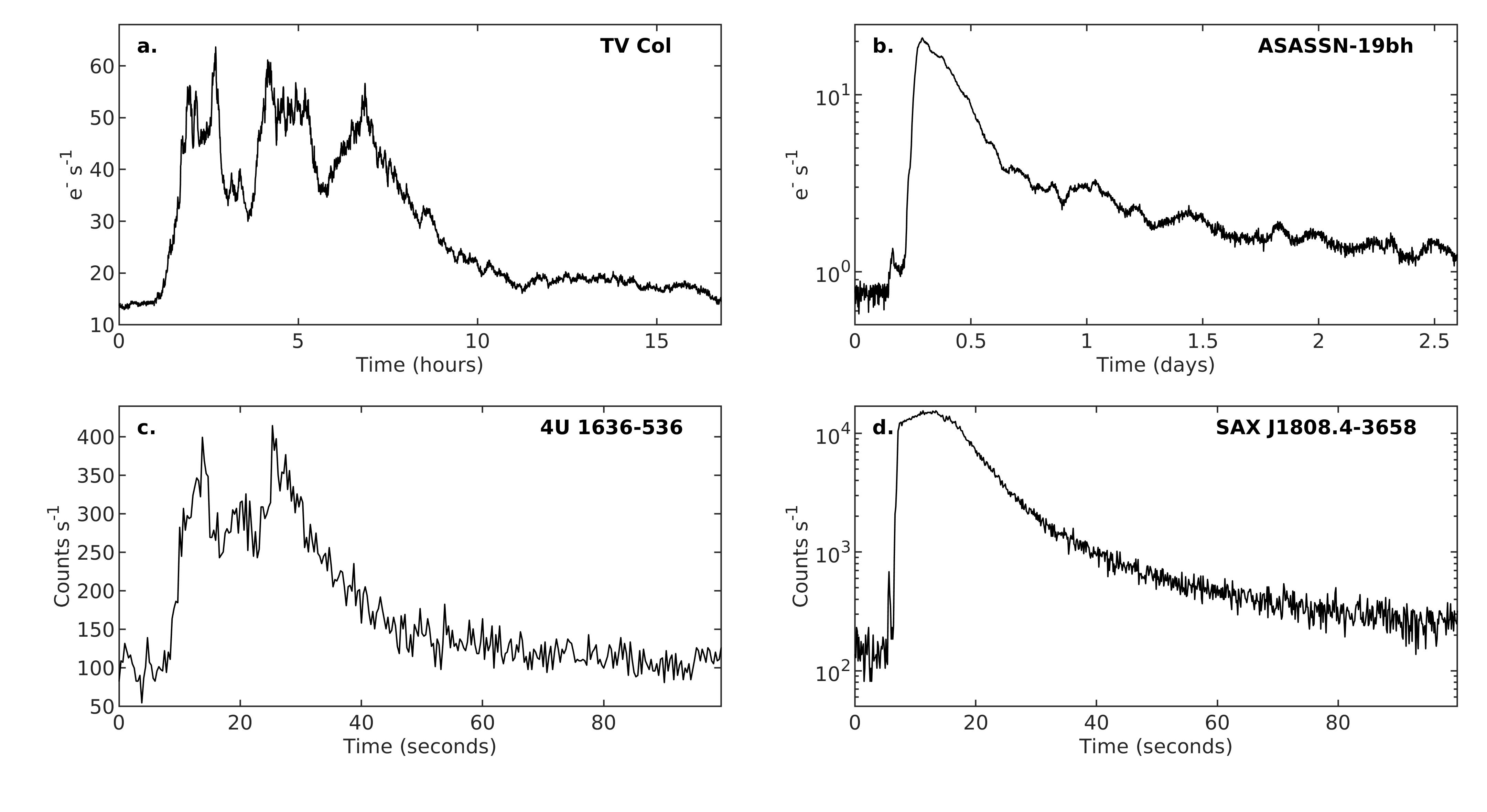}
\caption{
\noindent \textbf{Comparison between Type-I X-ray bursts and micronovae.}
\textbf{a.} \tess\ lightcurve of one of the rapid bursts observed in TV Col. \textbf{b.} \tess\ lightcurve of the rapid burst observed in ASASSN-19bh. \textbf{c.} \textit{EXOSAT}-ME X-ray lightcurve of 4U 1636$-$536 of one of the Type-I X-ray bursts. Note the similar multi-peak structure in both TV Col and 4U 1636$-$536. \textbf{d.} \textit{RXTE}-PCA X-ray lightcurve of one rapid burst in SAX J1808.4$-$3658. Note the precursor present in both ASASSN-19bh and SAX J1808.4$-$3658. In all panels the time axis has been arbitrarily shifted.
}
\end{center}
\end{extFigure*}

\begin{extFigure*}[ht]
\begin{center}
\includegraphics[width = \textwidth]{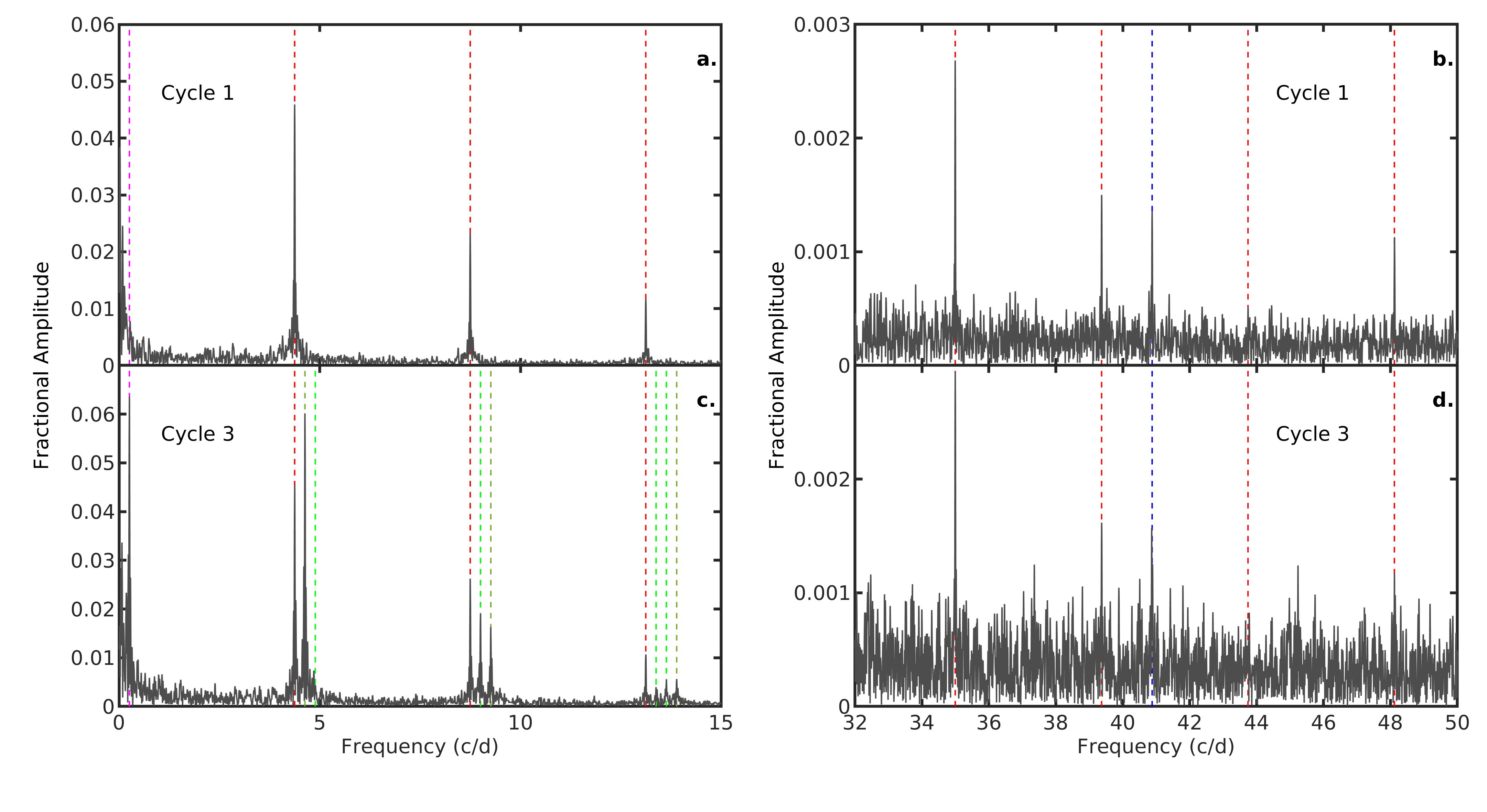}
\caption{
\noindent \textbf{Lomb-Scargle periodograms of TV Col.}
Periodograms using \tess\ data for TV Col \textbf{a.} Cycle 1 (120-s) low frequency periodogram. \textbf{b.} Cycle 1 high frequency periodogram. \textbf{c.} Cycle 3 (20-s) low frequency periodogram. \textbf{d.} Cycle 3 high frequency periodogram. In all panels the dashed-red lines mark the orbital frequency and associated harmonics, while the dashed-blue vertical lines mark the spin-to-orbital beat frequency. The dashed-magenta line marks the superorbital signal. The dashed-green lines mark the detected negative superhump, associated harmonics, and beats with the orbital frequency. 
}
\end{center}
\end{extFigure*}

\begin{extFigure*}[ht]
\begin{center}
\includegraphics[width = \textwidth]{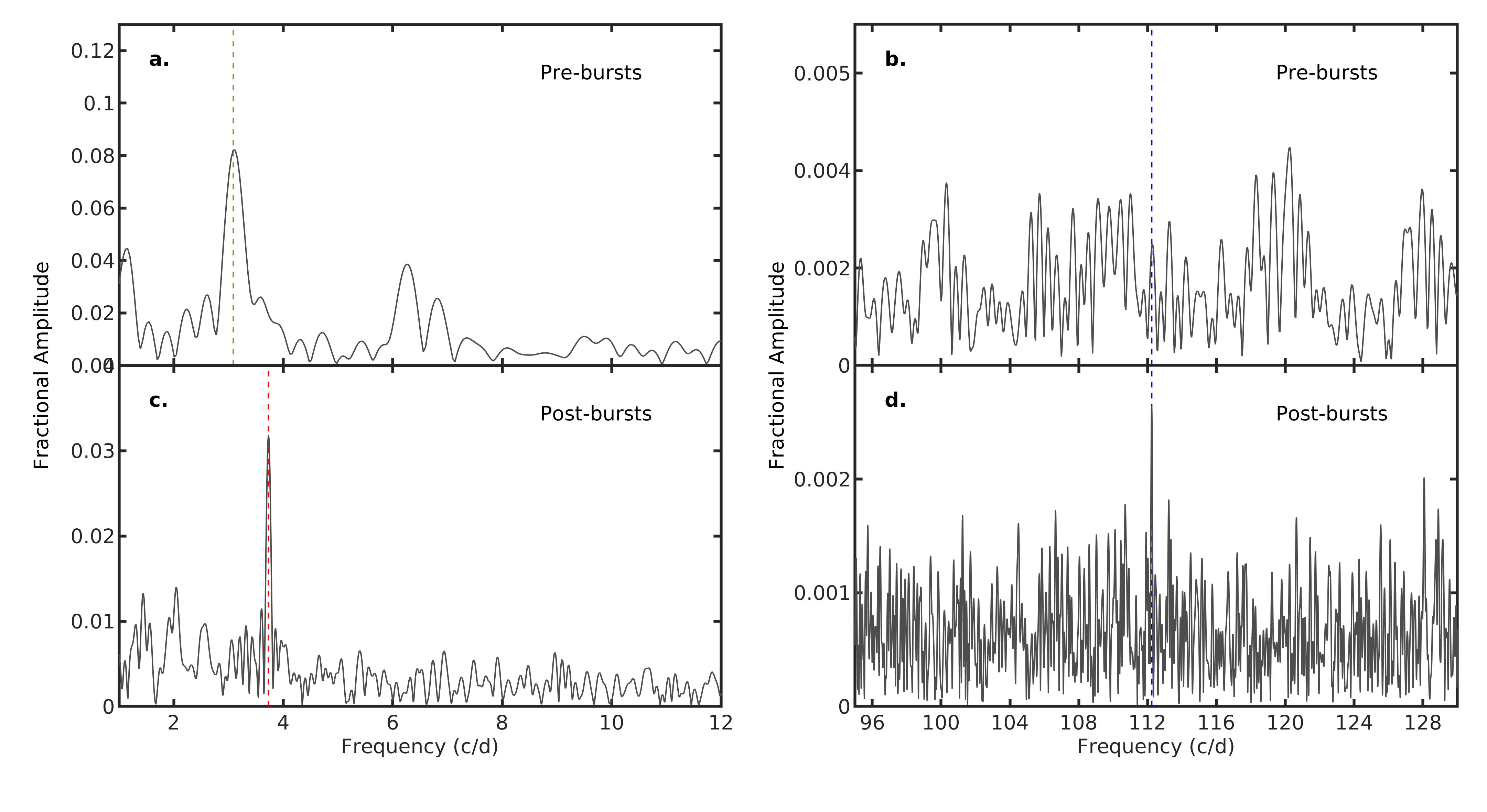}
\caption{
\noindent \textbf{Lomb-Scargle periodograms of EI UMa.}
Periodograms using TESS data for EI UMa during Cycle 2 (120-s cadence). \textbf{a.} Low frequency periodogram computed before the bursts. \textbf{b.} High-frequency periodogram computed before the bursts. \textbf{c.} Low-frequency periodogram computed after the bursts. \textbf{d.} High-frequency periodogram computed after the bursts. In all panels the dashed-red vertical line marks the detected orbital frequency. The dashed-blue vertical lines mark the spin-to-orbital beat frequency. The dashed-green line marks the positive superhump frequency.
}
\end{center}
\end{extFigure*}

\end{document}